On the Existence of the Kolmogorov Inertial Range in the Terrestrial Magnetosheath Turbulence

S. Y. Huang[1,2], L. Z. Hadid[2], F. Sahraoui[2], Z. G. Yuan[1], X. H. Deng[3]

[1] School of Electronic Information, Wuhan University, Wuhan, China

[2] Laboratoire de Physique des Plasmas, CNRS-Ecole Polytechnique-UPMC, Palaiseau, France

[3] Institute of Space Science and Technology, Nanchang University, Nanchang, China

**Abstract**

In the solar wind, power spectral density (PSD) of the magnetic field fluctuations generally follow the so-called Kolmogorov spectrum $f^{-5/3}$ in the inertial range, where the dynamics is thought to be dominated by nonlinear interactions between counter-propagating incompressible Alfvén wave parquets. These features are thought to be ubiquitous in space plasmas. The present study gives a new and more complex picture of magnetohydrodynamics (MHD) turbulence as observed in the terrestrial magnetosheath. The study uses three years of in-situ data from the Cluster mission to explore the nature of the magnetic fluctuations at MHD scales in different locations within the magnetosheath, including flanks and subsolar regions. It is found that the magnetic field fluctuations at MHD scales generally have a PSD close to $f^{-1}$ (shallower than the Kolmogorov one $f^{-5/3}$) down to the ion characteristic scale, which recalls the energy containing scales of solar wind turbulence. The Kolmogorov spectrum is observed only away from the bow shock toward the flank and the magnetopause regions in 17% of the analyzed time intervals. Measuring the magnetic compressibility, it is shown that only a fraction (35%) of the observed Kolmogorov spectra were populated by shear Alfvénic fluctuations, whereas the majority of the events (65%) was found to be dominated by compressible magnetosonic-like fluctuations, which contrasts with well-known turbulence properties in the solar wind. This study gives a first comprehensive view of the origin of the $f^{-1}$ and the transition to the Kolmogorov inertial range; both questions remain controversial in solar wind turbulence.

1. Introduction

Turbulence is ubiquitous in astrophysical plasmas such as the solar wind, planetary magnetospheres, the interstellar medium and accretion flows (Tu and Marsch 1995; Goldstein, 2001; Bruno and Carbone 2005; Sahraoui et al. 2006; Scheckochihin et al. 2009; Lazarian et al. 2012, Huang et al. 2012). Common to these diverse turbulent systems is the presence of an inertial range through which energy cascades from large to the small scales where dissipative mechanisms convert the turbulent energy into plasma heating. The near-Earth space provides an ideal laboratory to investigate plasma turbulence due to the wide range of temporal and spatial scales involved and to the availability of high quality *in-situ* measurements of the fields and the plasma particles. In the solar wind, the most studied astrophysical plasma using *in-situ* data, the magnetic energy spectra exhibit at least four dynamical ranges. First is the energy-containing range that has a scaling $\sim f^{-1}$ for frequencies $\leq 10^{-4}$ Hz (given in the spacecraft reference frame) (Bavassano et al. 1982). The origin of this range, also known as the energy driving scales (Bruno and Carbone 2005) or the "1/$f$ flicker noise" (Matthaeus & Goldstein, 1986), remains hotly debated. It includes the superposition of magnetic elements with different statistical properties that emerge from the corona due to magnetic reconnection

(Matthaeus & Goldstein, 1986; Matthaeus et al. 2007), evolution of the Alfvén waves coming from the corona and their reflection in the expanding solar wind (Velli et al. 1989; Verdini et al. 2012) and inverse cascade in MHD turbulence (Dmitruk & Matthaeus, 2007). It is worth recalling that the "1/f noise" spectrum is observed also in the solar photospheric magnetic field (Matthaeus et al. 2007) and in a variety of other systems that include electronic devices, dynamo experiments and geophysical flows (see Dmitruk et al. 2011, and the references therein). Second is the inertial range with a scaling $f^{-5/3}$ in the frequency range ~ $[10^{-4}, 10^{-1}]$ Hz. This range, where dissipation is assumed to be negligible (Kolmogorov, 1941), is thought to be generated via nonlinear interactions between counter-propagating incompressible Alfvén wave-packets (Iroshnikov 1963; Kraichnan 1965). Third is the range near the ion characteristic scales ~ [0.1, 1] Hz, known also as the dissipation range, where spectra can steepen significantly to ~ $f^{-4.5}$ (Goldstein et al. 1994; Leamon et al. 1998; Stawicki et al. 2001; Smith et al. 2006; Sahraoui et al. 2010; Bruno et al. 2014; He et al. 2015a). Fourth is the dispersive range far below the ion scale, ~ [3, 30] Hz, with a scaling ~ $f^{-\alpha}$ and α ∈ [-3.1, -2.3] (e.g., Alexandrova et al. 2012; Sahraoui et al. 2013). In this range, where dispersive and kinetic effects become important, the nature of the turbulent fluctuations is an unsettled question (Bale et al. 2005; Sahraoui et al. 2009, 2010; Podesta et al. 2012; Gary et al. 2013; He et al. 2011, 2012; Salem et al. 2012; Kiyani et al. 2013; Chen et al. 2013). A new steepening of the magnetic energy spectra has been reported near the electron scales ($f \geq 40$ Hz) and was interpreted as due to the ultimate dissipation of the residual magnetic energy into electron heating (Sahraoui et al. 2009). Due to instrumental limitations, the actual scaling of the magnetic energy spectra at sub-electron scales remains an open question in solar wind turbulence (Alexandrova et al. 2012; Sahraoui et al. 2013).

The magnetosheath, the region bounded by the bow-shock and the magnetopause, offers another alternative to study space plasma turbulence under more complex conditions than in the solar wind (e.g., Sahraoui et al. 2006; Karimabadi et al. 2014; Breuillard et al. 2016; Huang et al. 2016). The shocked solar wind exhibits indeed a variety of dynamical features such as heating and compression of the plasma, kinetic instabilities, particle beams and boundary effects due to the bow shock and the magnetopause (Sahraoui et al. 2004, 2006; Yordanova et al. 2008; Tsurutani and Stone, 1985). In a case study, Alexandrova et al. (2008) have shown the presence of a Kolmogorov spectrum $f^{-5/3}$ at MHD scales in the flank of terrestrial magnetosheath. Recent large 2D hybrid simulations showed a similar observation in the magnetosheath downstream of quasi-parallel shocks (Karimabadi et al. 2014). However, Hadid et al. (2015) performed a statistical survey of the Cassini data to investigate low frequency turbulence in the Saturn's magnetosheath, and found no evidence of the inertial range at MHD scales. Unlike the solar wind, the ubiquity of the Kolmogorov inertial range in magnetostheath turbulence has been therefore questioned. In this work, we use the Cluster data to explore the conditions of existence of the inertial range and to reveal the nature of terrestrial magnetosheath turbulence at MHD scales.

## 2. Results

### 2.1 Spectral slopes at MHD scales in the magnetosheath

We used three years (2001-2003) of the Cluster data covering a broad range of plasma parameters and different regions (sub-solar and flanks) of the terrestrial magnetosheath. The magnetic field data come from the Flux Gate Magnetometer (FGM) sampled at 23 Hz (Balogh et al. 1997) and the plasma data from the Cluster Ion Spectrometer (CIS) sampled each 4 seconds (Rème et al. 2001). We computed the PSD of the magnetic fluctuations for ~1600 of time intervals of ~1.6 hours. This time duration allows us to probe into frequencies as low as $10^{-4}$ Hz, which belong to the MHD scales (the typical ion gyro-period in the magnetosheath is of the order of 1s). Figure 1 shows examples of the spectra computed in the magnetosheath. They are characterized by two well-defined frequency bands exhibiting power-law like scaling and separated by a break occurring near the ion scales. The spectrum of Figure 1b shows a slope at MHD scales close to the Kolmogorov −5/3 scaling, whereas the one in Figure 1a is shallower and has a slope close to −1.

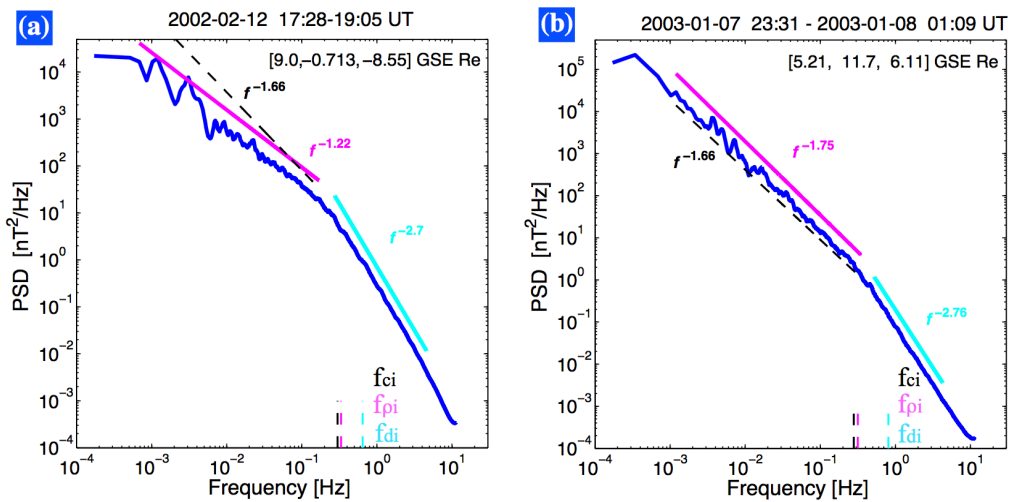

Figure 1. Two examples of the analyzed power spectral densities (PSDs) of magnetic fluctuations in the magnetosheath. The vertical black, green and magenta lines represent respectively the ion cyclotron frequency, the Taylor-shifted ion gyroradius scale and inertial length. The Kolmogorov scaling $f^{-5/3}$ is shown in black dashed lines for reference.

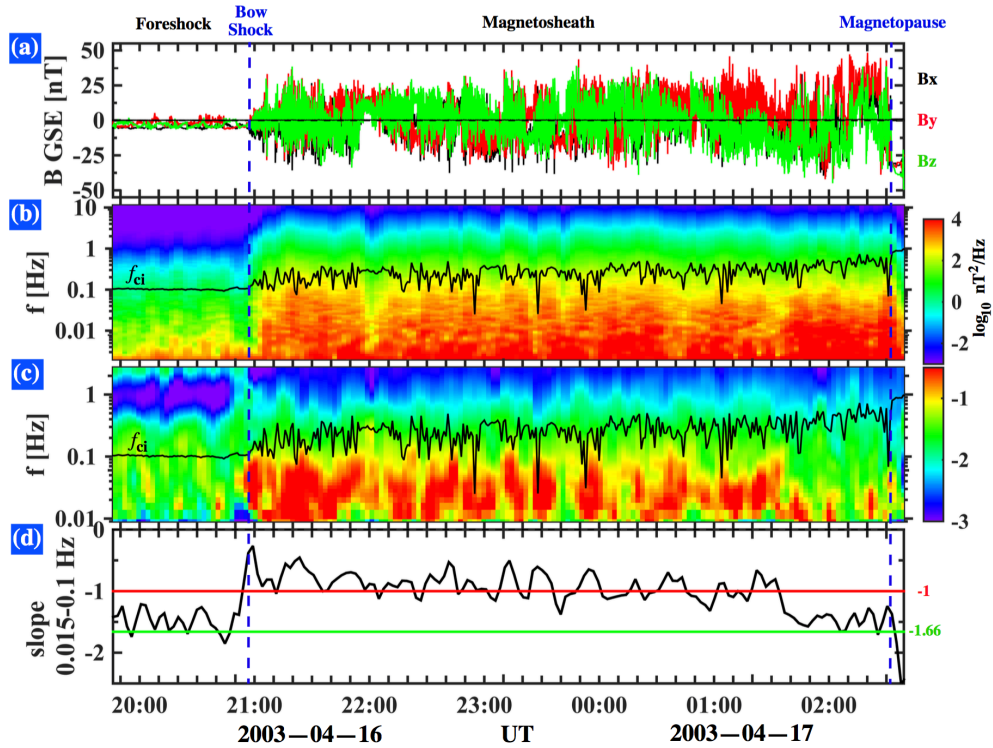

Figure 2. The time variations of the local slope during a full crossing of the magnetosheath. (a) The components of the magnetic field B showing a full crossing of the magnetosheath (from 21:00 UT on day 2003-04-16 to 02:00 UT on day 2003-04-17, (b) the corresponding spectrogram of the magnetic fluctuations, (c) the time-frequent-dependent spectral slopes of magnetic field, (d) the spectral slopes of magnetic field between 0.015 and 0.1 Hz indicating variation in the scaling of the turbulence at MHD scales as the spacecraft crosses different regions, from the solar wind/foreshock to the magnetopause.

To explore the origin of the difference in the scaling of the magnetic energy spectra, we display in Figure 2 a case study of a full magnetosheath crossing, from the solar wind/foreshock region (before 21:00 UT on day 2013-04-16) to the magnetopause (after 02:45 UT on day 2003-04-17). The local (time-dependent) spectral slope of δB shows a clear transition from ~ −5/3 to ~ −1 at the shock crossing then stays nearly constant for more than 4 hours before steepening to ~ −5/3 when approaching the magnetopause whose crossing occurs at about 01:45 UT. This observation shows that the scaling properties of solar wind turbulence do not "survive" after interacting with the bow shock; the new (dynamically young) fluctuations generated behind the shock (Karimabadi et al. 2014) were shown to be uncorrelated (i.e., their temporal increments had a quasi-Gaussian distribution) (Hadid et al. 2015).

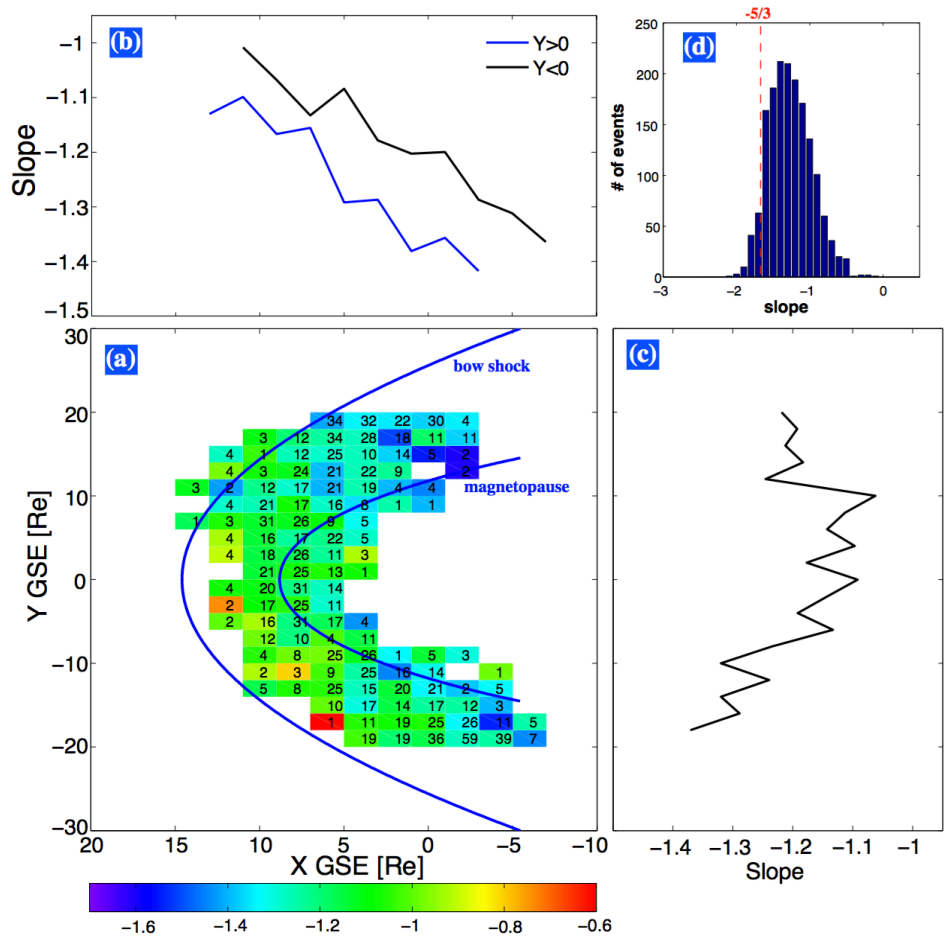

Figure 3: 2D distribution of the slopes $P_\alpha(X_{GSE}, Y_{GSE})$ at MHD scales. The 2D distribution $P_\alpha(X_{GSE}, Y_{GSE})$ of the spectral slopes at MHD (a), and the integrated ones along the $Y_{GSE}$ direction (b) and $X_{GSE}$ direction (c). The blue curves represent the average magnetopause and bow shock positions computed using the paraboloidal bow shock model of Filbert and Kellogg (1979) and the magnetopause model of Sibeck et al. (1991). The number in each bin

represents the number of events. (d) is the histogram of the measured spectral slopes where the red dashed line indicates the Kolmogorov slope -5/3.

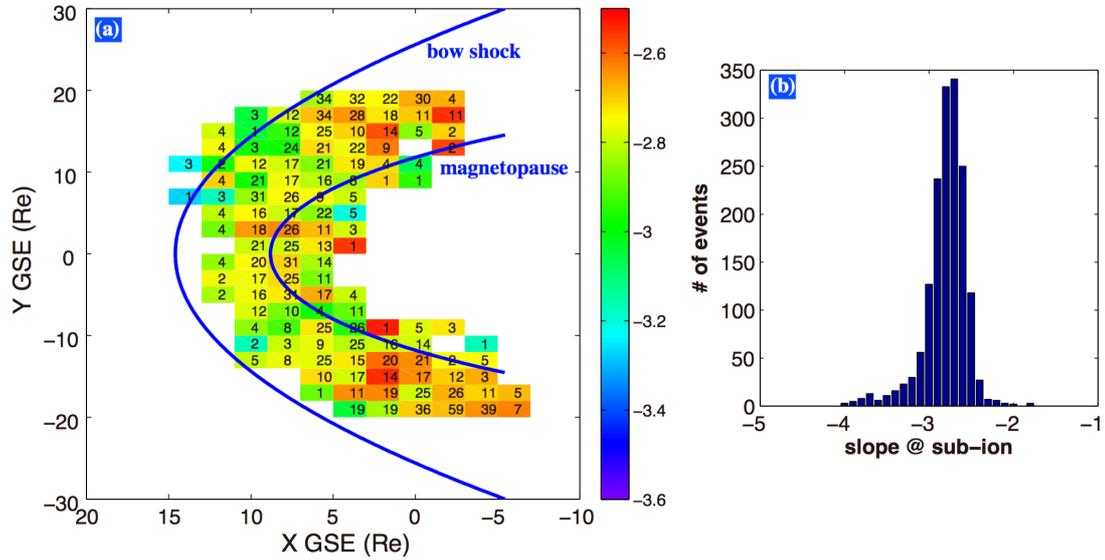

Figure 4: 2D distribution of the slopes $P_\alpha(X_{GSE}, Y_{GSE})$ at sub-ion scales. The 2D distribution $P_\alpha(X_{GSE}, Y_{GSE})$ of the spectral slopes at sub-ion scales (a), the histogram of the spectral slopes (b). The number in each bin represents the number of events.

To answer the question as to whether the previous results are statistically meaningful, we computed the spatial distribution of the spectral slopes $\alpha$ at MHD scales in different regions of the magnetosheath. The 2D distribution of the slope values $P_\alpha(X_{GSE}, Y_{GSE})$ as function of the location within the magnetosheath is displayed in Figure 3a (where GSE is the Geocentric Solar Ecliptic coordinates). One can see that the slopes ~ −1 were observed near the bow shock whereas steeper spectra close the Kolmogorov scaling $f^{-5/3}$ were observed closer to the magnetopause toward the flank regions. The same observation is made by examining the reduced 1D distribution of $P_\alpha$, i.e., integrated along $X_{GSE}$ (Figure 3c) then along $Y_{GSE}$ directions (Figure 3b). The distribution along $X_{GSE}$ is split into two parts, $Y_{GSE} > 0$ and $Y_{GSE} < 0$. It is clearly seen that the slopes decrease away from the bow shock (i.e., with decreasing $X_{GSE}$). The slopes along $Y_{GSE}$ flatten to −1 near the sub-solar region of the magnetosheath ($Y_{GSE} \sim 0$) and decrease slightly toward the flanks (i.e., with increasing $|Y_{GSE}|$). A study (not shown here) of the possible effect of the geometry of the shock did not demonstrate a significant correlation between the observed spectra and the angle $\theta_{nB}$ between the normal of the shock and the interplanetary magnetic field. The histogram shown in Figure 3d indicates that the spectral slopes vary within the interval ~ [−2.2, −0.3] with a peak near −1.2. The bulk of the distribution is significantly shallower than the Kolmogorov scaling $f^{-5/3}$ (red dashed line in Figure 3d) and that reported in the solar wind (e.g., Smith et al. 2006). These observations question the ubiquity of the Kolmogorov scaling in the terrestrial magnetosheath, although a small fraction (17%) of the events showed a Kolmogorov spectrum (here defined the spectral slope < -1.5).

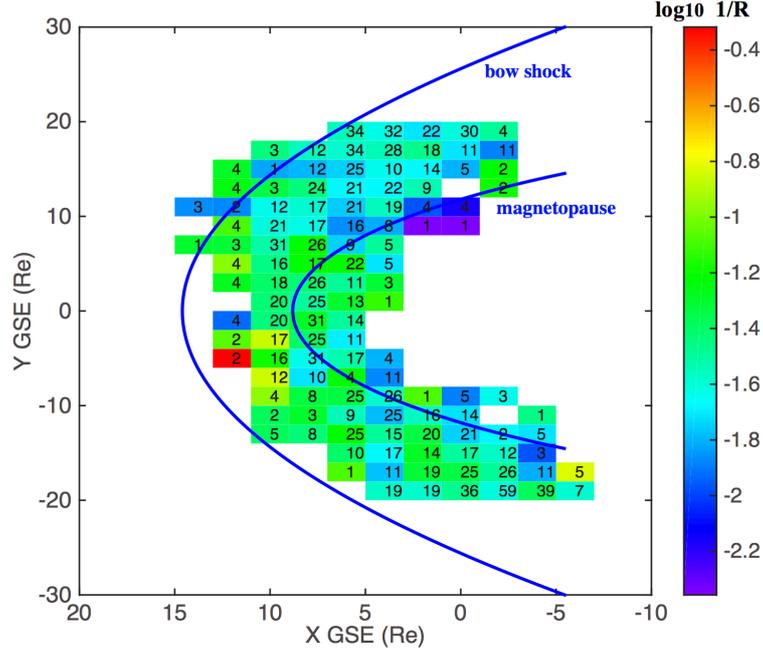

Figure 5: Distribution $P_{1/R}(X_{GSE}, Y_{GSE})$ in the magnetosheath for all statistical events. The ratio ($R = \tau_c f_b$) between the correlation time of the turbulent fluctuations $\tau_c = L_c/V_f$ and the local ion time scale which is considered here to be $f_b$ ($f_b$ is the frequency corresponding to the spectral break occurring near the ion scale). The number in each bin represents the number of events.

On the contrary, the slopes at the sub-ion scales were found not to depend on the location within the magnetosheath as can be seen in Figure 4a. This can be explained by the fact that the kinetic scales (≤100 km) are much smaller than the integral (size) scale of the magnetosheath (>$10^4$ km), meaning that the large scale boundaries do not influence the small (kinetic) scale turbulence. Furthermore, the histogram of the slopes (Figure 4b) is found to be similar to those reported in the solar wind (Alexandrova et al. 2012, Sahraoui et al. 2013) and in the magnetosheath (Huang et al. 2014, Hadid et al. 2015), with a peak near -2.8. This observation suggests that kinetic scale turbulence has a "universal-like" scaling distribution in the sense that it does not depend either on the explored region (magnetosheath or solar wind) or on the nature of the turbulence at MHD scale ($f^{-1}$ energy containing range or $f^{-5/3}$ inertial range), which is probably due to that kinetic scale turbulence is generated locally so its nature is not affected by large-scale inhomogeneities/fluctuations.

Hadid et al. (2015) showed that the $f^{-1}$ range in the magnetosheath of Saturn was populated by random-like fluctuations. It has been speculated that the interaction of the solar wind with the shock destroys the pre-existing correlations between the turbulent fluctuations, yielding the random-like nature of the fluctuations behind the shock. Nonlinear interactions between those fluctuations may lead to turbulent cascade away from the shock. In this scenario, as the distance increases from the shock, the cascade would involve larger and larger scales that deplete the energy content of the $f^{-1}$ range. Consequently, when the turbulence cascade is developed the correlation length of the turbulent fluctuation $L_c$ should be much larger than the characteristic ion scale in the magnetosheath $L_i$ (which can be associated to dissipation), i.e., $R = L_c/L_i \gg 1$. In other words, the separation of scales is needed to allow the turbulence cascade to proceed from the scales $\sim L_c$ to scales $\sim L_i$, leading to the emergence of the inertial range. In this case, the ration $R$ can be seen as a measure of the size of the inertial range. To test this hypothesis, we estimated $R$ as the ratio between the correlation time of the turbulent fluctuations $\tau_c$ and the local ion time scale $\tau_i$ (assuming the Taylor hypothesis $\tau \sim L/V_f$ where $V_f$ is the mean plasma velocity). Here the local ion scale is considered to be $\tau_i \sim 1/f_b$, where $f_b$ is the frequency corresponding to the spectral break occurring near the ion scale as those shown in Figure 1, meaning that $R = \tau_c f_b$. For a direct comparison with Figure 3(a), we plotted in Figure 5 the distribution $P_{1/R}(X_{GSE}, Y_{GSE})$ of the inverse of the estimated ratio $1/R$ for all events in a log scale. It shows a relatively good matching between the regions where the Kolmogorov scaling is observed (Figure 3a) and those where the correlation time of the turbulence is much larger than the (local) time scale of the ions. This observation is consistent with the scenario proposed in Hadid et al. (2015) in which fully developed turbulence may result from the intrinsic nonlinear evolution of the (dynamically young) random-like fluctuations generated behind the bow shock. This is supported by noting that the ratio $R$ is related to the effective magnetic Reynolds number $R_{eff}$ ($R_{eff} = R^2$) given in Weygand et al. (2007) if the Taylor micro-scale is considered here to be the ion (spectral break) scale. However, this scenario of a developing turbulence away from the shock does not rule out the possibility that the observed turbulence might have been generated as well by local sources such as the Kelvin-Helmholtz instability that develops on the flank of the magnetopause (e.g., Hasegawa et al. 2004; Karimabadi et al. 2014).

## 2.2 The nature of the Kolmogorov inertial range: Alfvénic *vs* compressible turbulence

In this section we focus on the nature of the turbulent fluctuations found to have a Kolmogorov scaling in the magnetosheath. To our knowledge, all Kolmogorov spectra reported in previous observations have been attributed to incompressible Alfvénic turbulence ($\delta B_\perp^2 >> \delta B_\parallel^2$) both in the solar wind (e.g., Goldstein et al. 1995; Sahraoui et al. 2010; Podesta et al. 2012; Kiyani et al. 2013) and in the magnetosheath (Alexandrova et al. 2008). Here, we give evidence of the existence of a turbulence dominated by compressible (i.e., non-Alfvénic) fluctuations which have $\delta B_\parallel^2 >> \delta B_\perp^2$ and a Kolmogorov spectrum $\sim f^{-5/3}$. To this end, we use the magnetic compressibility $C_\parallel$ given by the ratio between the PSDs of the parallel (w.r.t. the background field $\mathbf{B}_0$) to the total magnetic field fluctuation

$$C_\parallel(f) = |\delta B_\parallel(f)|^2 / (|\delta B_\parallel(f)|^2 + |\delta B_\perp(f)|^2) \qquad (1)$$

An example of theoretical magnetic compressibility calculated from the linear solutions of the Vlasov-Maxwell using the WHAMP code (Ronmark 1982) is shown in Figure 6a for typical magnetosheath plasma parameters. As one can see, the magnetic compressibility can allow one to clearly distinguish between Alfvénic and (fast or slow) magnetosonic turbulence at the large (MHD) scales (e.g., Gary et al. 2009; Sahraoui et al. 2012). Note that a parametric study based on the compressible Hall–MHD model (Sahraoui et al. 2003) (not shown here) showed that the magnetic compressibility of the three modes keep the same profile (but change its magnitude) when varying β in the range [0.2, 100] for a fixed $\theta_{kB}$ = 87° ($\theta_{kB}$ is the angle between wave vector and the ambient magnetic field) and when varying $\theta_{kB}$ from quasi-parallel to quasi-perpendicular angles for β = 1. In the present study we computed the magnetic compressibility for all the time intervals whose spectra showed a Kolmogorov-like scaling. Owing to the fact that the magnetic field components in the magnetosheath can be subject to large variations, we decompose the magnetic field fluctuations into the parallel and perpendicular directions using both the global mean field, computed as the average over each time interval of 1.6 hour, and the local (scale dependent) mean field computed using the wavelet technique described in Kiyani et al. (2013).

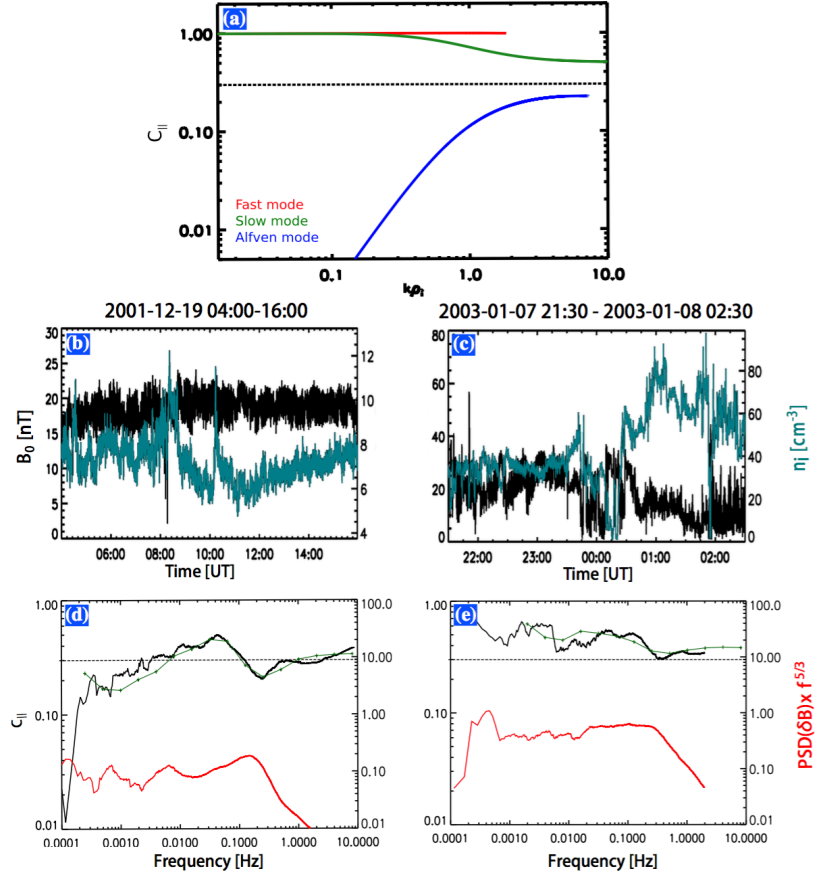

Figure 6. Theoretical magnetic compressibilities computed from the linear solutions and magnetic field and density measurements with the corresponding magnetic compressibilities of two different cases in the magnetosheath. (a) The red, green, and blue curves correspond respectively to the fast, slow, and Alfvén modes from the linear solutions of the Vlasov–Maxwell equations using the WHAMP code (Ronnmark, 1982) for $\beta = 1$ and $\theta_{kB} = 87°$. The horizontal dashed black line at $C_\parallel = 1/3$ indicates the power isotropy level. (b-c) Magnetic field magnitude and plasma density measured by the FGM and the CIS experiments onboard Cluster, (d-e) the magnetic compressibility $C_\parallel$ (defined in the text) using the global (black) and the local (green) mean field, and their PSDs of magnetic field multiplied by $f^{5/3}$.

Figure 6b-6e shows the results of that analysis. First, we observe a relatively uniform magnetic field magnitude in Figure 6b characteristic of incompressible Alfvénic fluctuations as frequently reported in the solar wind (e.g., Kiyani et al. 2009; Matteini et al. 2015). This contrasts with the case in Figure 6c where strong variations in the amplitude of magnetic field can be seen, which is generally anti-correlated with the density fluctuations. This suggests the dominance of slow-like magnetosonic turbulence (Klein et al. 2012; Yao et al. 2013; Hadid et al. 2015; He et al. 2015b). This suggestion is confirmed by the observed magnetic compressibility shown in Figure 6d-6e. One can see that for the case of Figure 6b the

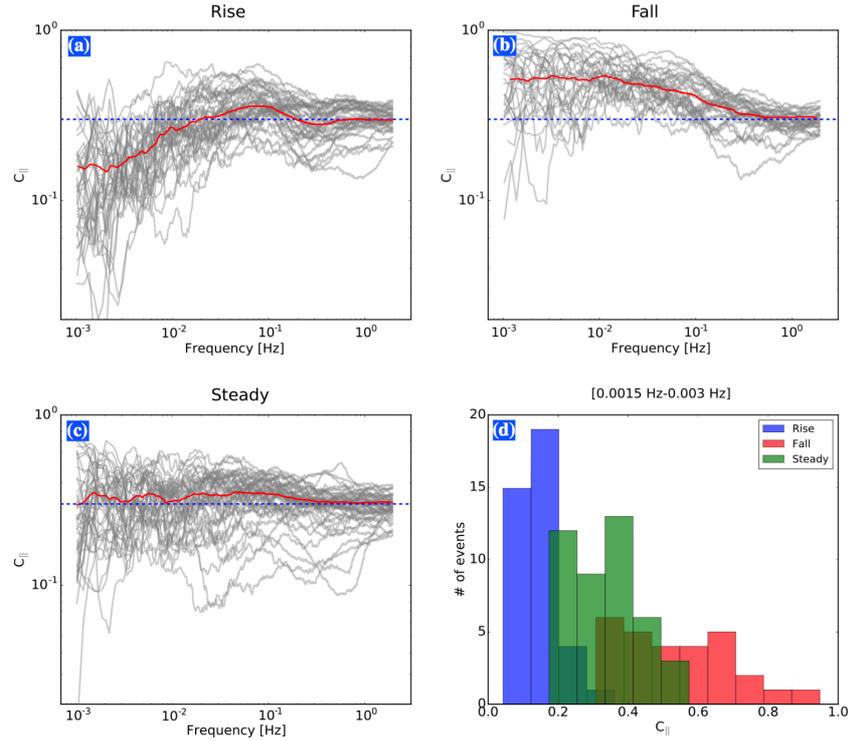

Figure 7: Estimated magnetic compressibility $C_\parallel$ for all statistical events which have a Kolmogorov-like scaling. Three distinct profiles were (grey curves: all events, red curves: mean values): (a) rising characteristic of shear Alfvén wave turbulence, (b) falling-off and (c) steady, both characteristic of compressible magnetosonic-like dominated turbulence. The dashed blues in indicate the value 1/3 of the compressibility. (d) shows the histogram of the averaged values of $C_\parallel$ in the indicated frequency range from 0.004 to 01 Hz.

observed magnetic compressibility resembles that of the Alfvén mode, whereas the one in Figure 6b is better fit by the slow or the fast magnetosonic ones as shown in Figure 6a. Note that the measured magnetic compressibility using the local and the global field decompositions give very similar results. According to the distinct trends of magnetic compressibilities for three different modes in Figure 6a, the statistical results are divided into three groups. An estimation of the magnetic compressibility for all the events that showed a

Kolmogorov-like spectrum $f^{-\alpha}$ with $1.5 < \alpha < 1.9$ is depicted in Figure 7a-7c (the red curves are the resulting mean magnetic compressibility for each panel). Three distinct profiles were evidenced: rising (a), falling (b) and steady (c) tones, based on the integrated value of the magnetic compressibility in the frequency bandwidth 0.00015-0.003Hz. These profiles reflect different properties of the turbulence: the rising profile is better reproduced by the Alfvén mode, while those of the falling and steady profiles fit better with the slow and fast magnetosonic modes, or a combination of both, as can be seen in Figure 6a. The relatively low magnetic compressibility of the steady profile in comparison with the falling one may as well be due to the presence of a small fraction of Alfvénic fluctuations.

## 3. Discussion

In this large statistical study we reported novel results about plasma turbulence in the terrestrial magnetosheath. First, we demonstrated that the turbulence properties at MHD scale, as known in the solar wind, do not survive after the interaction with the Earth's bow shock. Random-like fluctuations are generated behind the bow shock, leading to the formation of the $f^{-1}$ spectrum in the low frequency (MHD) range. We showed that these fluctuations can reach a fully developed turbulent state, characterized by a Kolmogorov spectrum $f^{-5/3}$, which has a correlation time much larger than the local ion time scale. This scenario is consistent with the observations of the Kolmogorov spectrum away from the bow shock toward the flanks of the magnetopause. This result questions the ubiquity of the Kolmogorov spectrum in space plasmas. It also gives a more comprehensive origin of the $f^{-1}$ spectrum and new insight into the transition of turbulence into the inertial range, knowing that both issues remain controversial in the solar wind. The second important result reported here is that, even for the events where the Kolmogorov spectrum is observed, the actual nature of the turbulence is very different from that in the solar wind. Most of the events (65%) were found to be dominated by the compressible magnetosonic-like modes. This observation underlines the need for developing a new phenomenology of compressible MHD turbulence different from the Iroshnikov-Kraichnan one used as the building block in incompressible MHD turbulence theories. Although great theoretical and numerical efforts have been done to investigate compressible MHD turbulence (e.g., Cho and Lazarian, 2002; Banerjee and Galtier, 2013), properties of magnetosheath turbulence as reported here require further theoretical investigation to be fully understood. Finally, this work can help improving current models of astrophysical turbulence by incorporating the compressible features observed here in the theoretical models used to study turbulence generated behind astrophysical shocks, in the heliosheath, in the interstellar medium and in supernova remnants (Scheckochihin et al. 2009. Schmidt et al. 2013; Vázquez-Semadeni et al. 1996), and possibly to revisit the studies of the origin and nature of the $f^{-1}$ spectrum in the solar wind.

This work was supported by the National Natural Science Foundation of China (41374168, 41404132, 41674161), Program for New Century Excellent Talents in University (NCET-13-0446), and China Postdoctoral Science Foundation Funded Project (2015T80830) and by LABEX Plas@Par through a grant managed by the Agence Nationale de la Recherche (ANR), as part of the program Investissements dAvenir under the reference ANR-11-IDEX-0004-02. SYH, LZH and FS acknowledge financial support from the project

THESOW, grant ANR-11-JS56-0008. L. H. acknowledges the use of the local field decomposition routine developed by K. Kiyani. S.Y.H acknowledges the useful discussions with Dr. H. Breuillard. The data come from the ESA/CSA and the AMDA websites. The French involvement in the Cluster mission is supported by CNES and CNRS.